\documentclass{amsart}

\usepackage{amsmath}
\usepackage{amssymb}
\usepackage{color}
\usepackage{amsxtra} 
\usepackage{mathrsfs} 
\usepackage{yfonts} 
\usepackage{hyperref}
\usepackage{bm}
\usepackage{graphicx}
\numberwithin{equation}{section}

\newtheorem{theorem}{Theorem}[section]

\newtheorem{thm}{Theorem} 
\theoremstyle{definition}
\newtheorem{definition}{Definition}

\theoremstyle{remark}
\newtheorem{remark}{Remark}

\newtheorem{observation}[theorem]{Observation}

   \makeatletter
\renewcommand*{\eqref}[1]{%
  \hyperref[{#1}]{\textup{\tagform@{\ref*{#1}}}}%
}
\makeatother

\allowdisplaybreaks[3]

\newcommand{\bkt}[1]{\langle#1\rangle}

\newcommand{\cal}{\mathcal}

\newcommand{\Bo}[1]{\mathbb{#1}}

\newcommand{\norm}[2]{\big\| #1 \big\| _{#2}}

\newcommand{\eqq}[1]{\begin{align*} #1 \end{align*}}

\begin{document}

\title[Reformulation of Kolmogorov-Richardson energy cascade
]
{Mathematical reformulation of the
    Kolmogorov-Richardson energy cascade in terms of vortex
    stretching
} 
\author{Tsuyoshi Yoneda} 
\address{Graduate School of Economics, Hitotsubashi University, 2-1 Naka, Kunitachi, Tokyo 186-8601, Japan} 
\email{t.yoneda@r.hit-u.ac.jp} 

\author{Susumu Goto} 
\address{Graduate School of Engineering Science, Osaka University, 1-3 Machikaneyama, Toyonaka, Osaka 560-8531, Japan} 
\email{s.goto.es@osaka-u.ac.jp}


\author{Tomonori Tsuruhashi} 
\address{Graduate School of Mathematical Sciences, University of Tokyo, Komaba 3-8-1 Meguro, 
Tokyo 153-8914, Japan} 
\email{tomonori@ms.u-tokyo.ac.jp}


\date{\today} 



\begin{abstract}
In this paper, with the aid of direct numerical simulations (DNS) of forced turbulence in a periodic domain, we mathematically reformulate the 
Kolmogorov-Richardson energy cascade in terms of vortex stretching. By using the description, we prove that if the Navier-Stokes flow satisfies a new regularity  criterion in terms of the enstrophy production rate, then the flow does not blow up. Our DNS results seem to support this regularity criterion. Next, we mathematically construct the hierarchy of tubular vortices, which is statistically self-similar in the inertial range. Under the assumptions of the scale-locally of the vortex stretching/compressing (i.e.~energy cascade) process and the statistical independence between vortices that are not directly stretched or compressed, we can derive the $-5/3$ power law of the energy spectrum of statistically stationary turbulence without directly using the Kolmogorov hypotheses.
\end{abstract} 

\maketitle


\section{Introduction}

Since the seminal study by Orszag and Patterson (1972)
\cite{Orszag-1972}, the numerical integration (i.e.~the direct
numerical simulations, DNS) of the Navier-Stokes equation without any
turbulence model has been playing important roles in turbulence
researches; see Ref.~\cite{Ishihara-2009} for a review of DNS of
turbulence. One of the most important features unveiled by
DNS of turbulence is that turbulence is not random but composed of
coherent motions. In particular, recent DNS
\cite{Goto-2008,Goto-2017,Motoori-2019,Motoori-2021} of
turbulence at sufficiently high Reynolds numbers have revealed that
there exists a hierarchy of coherent vortices in developed turbulence.

In the present study, we investigate the energy cascade
picture \cite{Tennekes,Frisch-1991} in turbulence of an
incompressible fluid with uniform density and kinematic viscosity $\nu$
under periodic boundary conditions in three orthogonal directions; and, 
on the basis of the picture, we develop arguments on the global-in-time
smooth solution of the Navier-Stokes equation. Since we need external
force to drive flow in a periodic cube, we impose
body force $f$ which injects the kinetic energy at a large scale 
$\cal L$.  Here, we define the Reynolds number by
\begin{equation}
 Re=U{\cal L}/\nu
\end{equation}
with $U$ being the characteristic velocity directly driven by $f$.
When $Re$ is sufficiently high (more concretely, higher than $O(10^4)$
\cite{Dimotakis-2000}), turbulence is composed of vortices with various
length scales. We reemphasize that these vortices are coherent in time
and space. The hierarchy of coherent vortices and its sustaining
mechanism of this statistically stationary turbulence in a periodic cube
are rather simple \cite{Goto-2017}. If we visualize vortices by using
the magnitude of vortices, we can only observe the forest of the
smallest-scale vortices (Fig.~1 in Ref.~\cite{Ishihara-2009}). This is
because the vorticity is predominantly determined by the smallest-scale
eddies. Therefore, we need a scale decomposition to capture the
hierarchy of coherent vortices. More
precisely, the band-pass filter (i.e.~the Littlewood-Paley decomposition)
of the Fourier components of the vorticity reveals a clear hierarchical
structures of coherent tubular vortices (see Fig.~2 in
Ref.~\cite{Goto-2017}). The sustaining mechanism of the hierarchy of
vortices is also simple. Tubular vortices at a given length (i.e.~a
given thickness) are stretched and amplified in straining motions around
(pairs of) larger tubular vortices. Here, it is important that this
generation process of smaller vortices by larger ones occurs locally in
scale. More quantitatively, 2--8 times larger vortices stretch
and generate vortices at a given scale (see Sec.~\ref{sec:transfer} for the present DNS
results). In other words, this sustaining process of turbulence is
consistent with the classical Kolmogorov theory (i.e.~the
Kolmogorov-Richardson energy cascade \cite{Frisch-1991}). It is also important
that the smallest length scale, namely the Kolmogorov length $\eta$, is
determined by the length for which the two time scales of the vortex
stretching and viscous effects are balanced. Therefore, the hierarchy
of vortices is sustained in the inertial range of the length scales
between $\cal L$ and $\eta$, but vortices cannot be amplified in the
viscous dissipation range of the length scales smaller than $\eta$.


The above-mentioned scale locality of the vortex stretching is also
consistent with the fact that the strain-rate, $U/{\cal L}$, induced by
the largest scale, $\cal L$, flow cannot overcome the viscous effects 
for the length scales smaller than the Taylor length $\lambda$
\footnote{We consider statistically stationary turbulence
and we define length scales ($\cal L$, $\lambda$ and $\eta$) by their
 averages. By the definition of the Taylor length and the
dissipation law \cite{Taylor-1935I} for statistically stationary
turbulence we obtain $\nu U^2/\lambda^2\sim\epsilon\sim U^3/{\cal L}$,
which reduces to the balance of the two time scales:
$\lambda^2/\nu\sim {\cal L}/U$. 
Although it is an important issue to investigate the temporal
fluctuations of the energy dissipation rate and these length scales
\cite{Vassilicos-2015,Goto-2016}, this is out of scope of the 
present study. Here, we investigate the averaged picture of
the energy cascade
in a sufficiently large domain.}.
Note that $\lambda\gg\eta$ for high Reynolds numbers. It is also
important that the ratio ${\cal L}/\eta$ increases with the 
Reynolds number as ${\cal L}/\eta\sim Re^{3/4}$. Hence, physically 
speaking:

\begin{observation}\label{physical observation}
The increase of $Re$ with fixing $f$ (i.e.~the decrease of
$\nu$ with fixing $U$ and $\cal L$) simply adds the number of the
levels of the hierarchy of vortices.
\end{observation}

\vspace{0.3cm}

In other words, this physical
picture may allow us to expect that a finite energy dissipation rate
$\epsilon$ can be sustained in the shear flow around thin coherent
tubular vortices even in the limit $\nu\rightarrow0$.
\begin{remark}
Jeong and the first author \cite{JY} mathematically considered the balance  between the
viscous dissipation and the vortex stretching.
More precisely, they prepared small-scale vortex blob and large-scale anti-parallel vortex tubes for the initial data (in other words, shear flows around thin coherent
tubular vortices), and showed that the corresponding 3D Navier–Stokes flow creates
instantaneous vortex-stretching. Using this stretching, they showed that the
flows satisfy a modified version of the zeroth law (but very close to the actual one) in a uniform time interval which in particular implies enhanced dissipation.
The zeroth law postulates that, under the normalization
of the initial data (depending on $\nu$) $\|u_{0}^\nu\|_{L^2}=1$,
the mean energy dissipation rate of the corresponding 3D Navier-Stokes flow $u^\nu$ does not vanish as $\nu\to 0$:
\begin{equation*}
\liminf_{\nu\to 0}\nu\langle\!\langle|\nabla u^\nu|^2\rangle\!\rangle>0,
\end{equation*}
where $\langle\!\langle\cdot\rangle\!\rangle$ denotes some ensemble or long-time, space averages.
\end{remark} 

The main purpose of the present study is to mathematically describe the
feature of solutions of the Navier-Stokes equation on the basis of these
physical pictures of turbulence. More concretely, with the aid of the
DNS of turbulence in a periodic cube (the details of which are given in
Sec.~\ref{s:DNS}),  first, we observe that the Navier-Stokes turbulence is mathematically smooth enough (Sec.~\ref{regularity-criterion}).
Second, we propose a mathematical
description of the energy cascading process in developed turbulence (Sec.~\ref{sophisticated}).
Then, we derive the  Kolmogorov $-5/3$ power law of the energy spectrum by employing this energy cascade picture, without directly using the Kolmogorov hypotheses. (Sec.~\ref{derivation of K41}). 

\vspace{0.4cm}

\noindent
{\bf Notation.}\quad
We use $A \lesssim B$ (equivalently, $B\gtrsim A$) if there is an absolute constant $C>0$ such that $A \le CB$. Then, we say $A \approx B$ if $A \lesssim B$ and $B\lesssim A$. Moreover, $A \lesssim_\gamma B$ means  $A \lesssim C_\gamma B$
with some constant $C_\gamma$ depending on $\gamma$. 

\section{Numerical simulations}
\label{s:DNS}

In this section, we summarize the numerical method and parameters, and
we show the results of our DNS.

\subsection{Numerical method}

We numerically integrate the Navier-Stokes equation in 
$\mathbb{T}^3:=(\mathbb{R}/2\pi\mathbb{Z})^3$:
\begin{equation}\label{NS}
  \left\{
   \begin{aligned}
    \partial_t u + u\cdot\nabla u+\nabla p = \nu \Delta u + f,\\
    \nabla\cdot u = 0,\\
    u(t=0) = u_0
   \end{aligned}
  \right.
\end{equation}
with the external force,
\begin{equation}
 f=( -\sin x \cos y, +\cos x \sin y, 0).
\end{equation}
Here, $u : [0,\infty)\times \mathbb T^3 \rightarrow \mathbb R^3$ and $p : [0,\infty)\times \mathbb
T^3 \rightarrow \mathbb R$ denote the velocity and pressure of the
fluid, respectively. We use the standard Fourier spectral method. Here, we numerically
integrate, by the fourth order Runge-Kutta-Gill scheme, the vorticity
equation in terms of the Fourier component $\widehat{\omega}(K)$ 
of vorticity, which is defined by
\begin{equation}
 \omega(x)
 =
 \sum_{K\in\Bo{Z}^3}
 \widehat{\omega}(K)e^{iK\cdot x}
 \quad
 \text{with}
 \quad
 ({\mathcal F}u)(K)
 = 
 \widehat{u}(K)
 :=
 \frac{1}{(2\pi)^{3}}\int_{\Bo{T}^3}u(x)e^{-iK\cdot x}\,dx.
\end{equation}
Note that we need to integrate only two components
of $\widehat{\omega}(K)$ because of the solenoidal condition
$K\cdot\widehat{\omega}=0$. For numerical efficiency, we use the fast
Fourier transform to evaluate the nonlinear term in the vorticity
equation in the real space. Here, we remove the aliasing errors 
by the phase shift method.

\subsection{Numerical parameters}

Since we fix the external force $f$, we change the Reynolds number
by changing $\nu$. We denote the spatial average of the energy
dissipation rate by $\epsilon(t)$. Then, the Kolmogorov length scale
is expressed by $\eta(t)=\epsilon(t)^{-1/4}\nu^{3/4}$. We choose the
number $N^3$ of the Fourier modes so that we can resolve the small-scale
structures of size $\eta(t)$. In the numerical simulations of 
turbulence, we use the indicator of the resolution 
$k_{\text{max}}\bkt{\eta}$, where $\bkt{\eta}$ is the 
temporal average of $\eta(t)$ and the maximum wavenumber 
$k_{\text{max}}=\sqrt{2}N/3$ is the present numerical scheme. Usually,
the condition $k_{\text{max}}\bkt{\eta}>1$ is recommended. In the 
present study, however, we adopt much larger value (say, $O(10)$; see
table \ref{t:para}) to resolve flow structures much smaller than 
$\bkt{\eta}$. We also examine that the small-scale statistics 
are independent of artificial parameters by changing $N^3$ with 
a common value of $\nu$.

We choose the time increment of the numerical integration so that the
CFL condition is satisfied.
We set the integration time of each run to be longer than the turnover time of the largest eddies.
We evaluate the development of turbulence by the Reynolds number
\begin{equation}
 R_\lambda(t)=u'(t)\lambda(t)/\nu
 \:,
\end{equation}
where $\lambda$ is the Taylor length and $u'(t)$ (i.e.~the standard
deviation of a velocity component) denotes turbulence intensity. 
In homogeneous isotropic turbulence, 
\begin{equation}
 R_\lambda(t)=\sqrt{\frac{20}{3\nu\epsilon(t)}}\:\mathcal{E}(t)
 \:,
\end{equation}
where $\mathcal{E}(t)$ is the kinetic energy per unit mass. In general, 
turbulence is fully developed when $R_\lambda>140$
\cite{Dimotakis-2000}. Turbulence with lower $R_\lambda$, the scale
separation between the largest scale, i.e.~the forcing scale, and $\eta$
is insufficient.

In the following, we show our numerical results obtained by the
simulation with the parameters given in table \ref{t:para}.

\begin{table}
\begin{center}
\begin{tabular*}{0.5\textwidth}{l|cccc}
Run &
$N^3$ &
$\nu$ &
$k_{\text{max}}\bkt{\eta}$ &
$\bkt{R_\lambda}$ 
\\\hline\hline
2-1K&
$1024^3$ &
$0.002$ &
 $3.0$ &
 $296$ \\\hline
8-1K&
$1024^3$ &
$0.008$ &
$9.3$ &
 $123$ \\\hline
8-2K&
$2048^3$ &
$0.008$ &
 $20$ &
 $136$ \\
\end{tabular*}
\end{center}
\caption{Numerical parameters. $N^3$, the number of the Fourier modes;
$\nu$, the kinematic viscosity; $\bkt{\eta}$, the temporal average of 
the Kolmogorov length scale ($k_\text{max}=\sqrt{2}N/3$ denotes the
maximum wavenumber); $\bkt{R_\lambda}$, the temporal average of the 
Taylor-length Reynolds number.}
\label{t:para}
\end{table}
%

\subsection{Energy transfer due to vortex stretching}
\label{sec:transfer}

\subsubsection{Four wavenumber ranges}

\begin{figure}
\begin{center}
\hspace{-2.2cm}
\begin{tabular}{cccc}
\includegraphics[bb=100 0 443 603,width=0.4\textwidth]{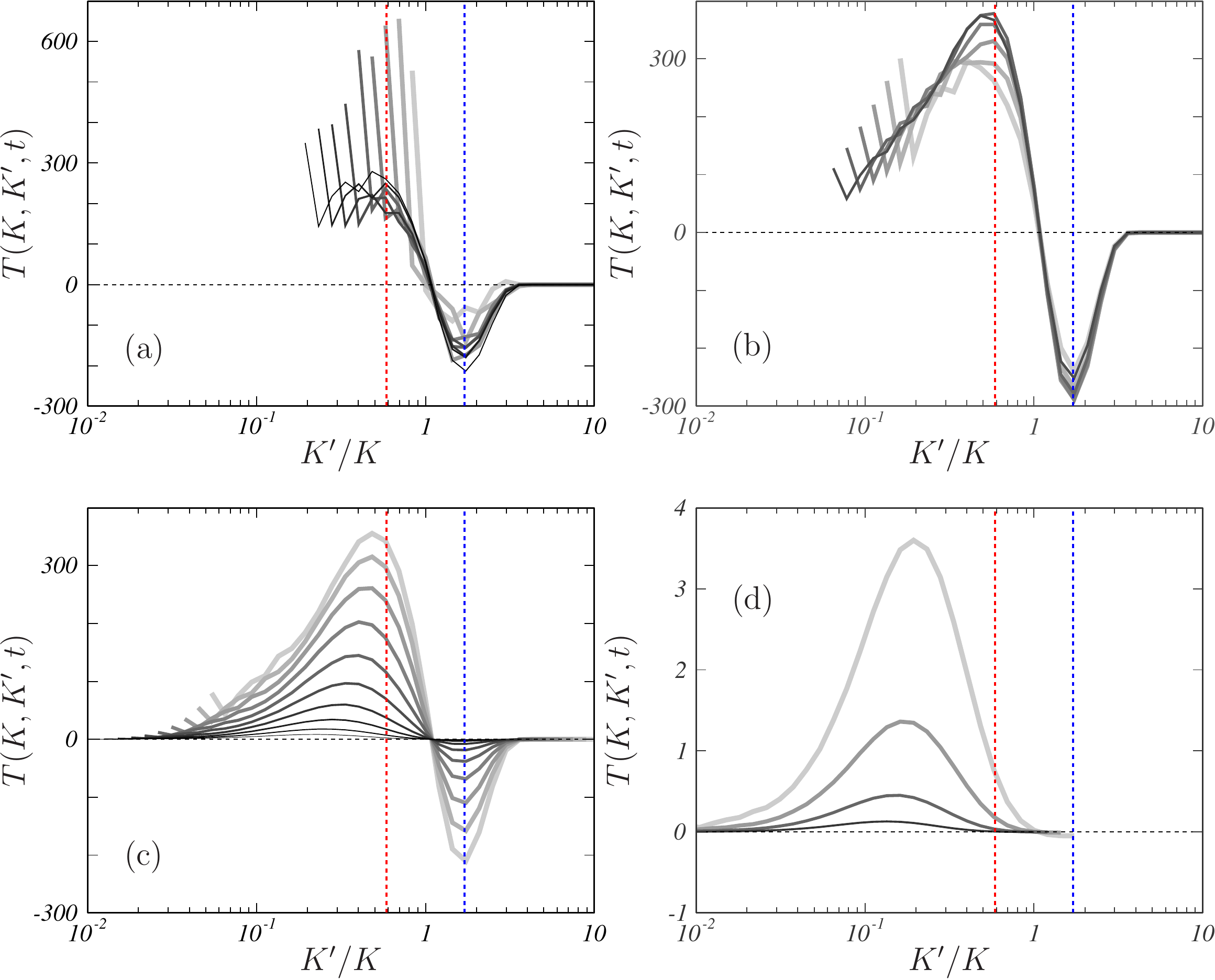}
\end{tabular}
\end{center}
\caption{Energy transfer $T(K,K',t)$ to the Fourier modes at wavenumber
$K$ from the Fourier modes at $K'$ due to vortex stretching
in the four different wavenumber ranges. In each plot, darker (thinner)
curves correspond to larger $K$. Red and blue vertical
lines indicate $K'=0.58K(=K/1.7)$ and $K'=1.7K$, respectively.
Note that, from this observation, we assume $\alpha=1.7$ in Sec.\!\!\! \ref{sophisticated}.
(a) In the energy containing range 
$K\in[1.2^{-34}\eta^{-1},1.2^{-26}\eta^{-1}]=[0.0020\eta^{-1},0.0087\eta^{-1}]$,
where the Fourier modes (which correspond to the whiskers
observed at the lower wavenumber) directly driven by external force.
(b) In the inertial range 
($K\in[1.2^{-25}\eta^{-1},1.2^{-20}\eta^{-1}]=[0.010\eta^{-1},0.026\eta^{-1}]$),
where we can observe a clear collapse of the curves for different $K$,
and the peaks are located at $K'=0.58K$ and the valleys are at $K'=1.7K$.
(c) In the dissipation range
$K\in[1.2^{-19}\eta^{-1},1.2^{-10}\eta^{-1}]=[0.031\eta^{-1},0.16\eta^{-1}]$,
where we do not observe the self-similarity of the energy transfer.
(d) In the far dissipation range
$K\in[1.2^{-9}\eta^{-1},1.2^{-6}\eta^{-1}]=[0.19\eta^{-1},0.33\eta^{-1}]$,
where the energy is dissipated by viscosity and almost no energy is
transferred to higher wavenumbers $|T(K'>K)|\ll|T(K>K')|$. 
Results of Run 2-1K ($R_\lambda\approx280$).}
\label{f:T}
\end{figure}

There are four distinct wavenumber ranges (or equivalently, scale
ranges) in each of which the energy transfer due to vortex stretching
exhibits different characteristics. To show this, we define
the energy transfer due to vortex stretching (see the energy equation \eqref{modern-energy-balance-0}) by
\begin{equation}\label{energy transfer due to vortex stretching}
 T(K,K')
 :=
 \frac{1}{K^2}\:
 \int_{\mathbb{T}^3}\omega(K)\cdot\nabla u(K')\cdot\omega(K)
 \:dx
 \:
\end{equation}
where
\begin{equation}\label{LP-decomposition-2}
 \omega(K)
 := 
 {\mathcal F}^{-1}\Big[\chi_{[K/\sqrt{2},\sqrt{2}K)}\:{\mathcal F}[\omega]\Big]
\end{equation}
and
\begin{equation}\label{bandpass-filter-K}
 u(K)
 := 
 {\cal F}^{-1}\Big[\chi_{[K/\sqrt{2},\sqrt{2}K)}\:{\cal F}[u]\Big]
 \:.
\end{equation}
Note that in the next section we will define $\omega^k$ by 
$\omega(2^k)$. We plot $T(K,K',t)$ in turbulence at
$R_\lambda\approx280$ (Run 2-1K) in figure \ref{f:T}. This figure
shows results at a single time. We have confirmed that the
behaviors of $T$ are qualitatively independent of time, though the values of $T$ evolve in
time.

First, let us look at the results shown in figure \ref{f:T}(a) for 
a low wavenumber range ($K\in[0.0020\eta^{-1}, 0.0087\eta^{-1}]$),
where the effect of the external force is dominant. This 
figure shows the energy transfer $T(K,K',t)$ as a function of 
$K'/K$, where the energy transfers from wavenumber $K'$ to $K$. 
The different curves in the plot show the results for different $K$;
darker (thinner) curves correspond to larger $K$. It is clear in this
figure that the Fourier modes at $K$ acquire energy from
smaller wavenumber modes ($K'<K$) and transfer the energy to larger 
wavenumber modes ($K<K'$). These characteristics are common in the four
wavenumber ranges shown in figures \ref{f:T}(a)--(d). However, we observe
in figure \ref{f:T}(a) ``whiskers'' at the lowest wavenumber. This 
corresponds to the contribution from the Fourier modes which the
external force $f$ directly drives. Recall that, in this turbulence,
the external force is steady and therefore these modes are rather
robust.

\begin{figure}
\begin{center}
\includegraphics[bb=0 0 429 293,width=0.8\textwidth]{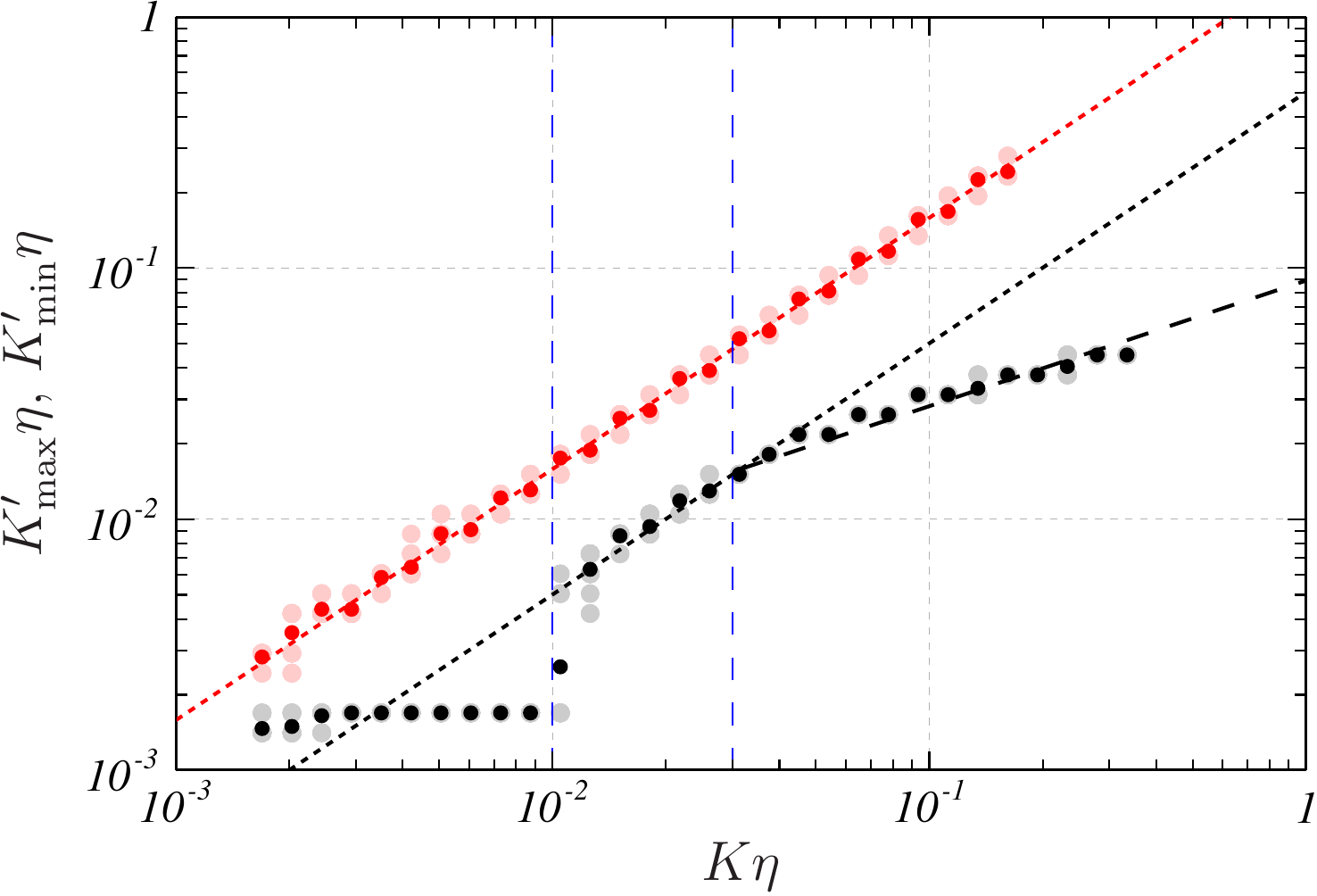}
\end{center}
\caption{The wavenumbers $K'_\text{max}$ (gray symbols) and
$K'_\text{min}$ (pink) giving the maximum and minimum of
$T(K,K',t)$, respectively, as a function of $K$. We plot the data
at 20 different instants. The red and black dots
show the temporal averages of $K'_\text{max}$ and $K'_\text{min}$,
respectively. The dotted and dashed lines indicate
$K'\sim K$ and $K'\sim K^{1/2}$, respectively. 
The two blue vertical dashed lines indicate $K=0.01\eta^{-1}$ and $0.03\eta^{-1}$.
Results of Run 2-1K 
($R_\lambda\approx280$). We do not show 
$K'_{\text{min}}$ for $K\eta>0.19$ because the numerical resolution is not
sufficient to determine it.} 
\label{f:scalings}
\end{figure}

Secondly, we look at the inertial range. Figure \ref{f:T}(b) shows the 
energy transfer in the wavenumber range
$K\in[0.0105\eta^{-1},0.026\eta^{-1}]$. This figure also shows
that the energy transfers from lower ($K'<K$) to higher $K'>K$ 
wavenumber modes. In contrast to the lower wavenumber range
[figure \ref{f:T}(a)], the contribution from about the half (more
precisely, $K'=0.58K$) wavenumber modes (i.e.~$1.7$ times larger 
scales) transfer the energy to $K$ than the forcing modes. Note
that the peaks at $K'=0.58K$ is higher than the whiskers. This is also
consistent with the observation in the same figure
[figure \ref{f:T}(b)] that the energy at the wavenumber $K$ transfers
most to a higher wavenumber mode at $K'=1.7K$ (i.e.~$0.58$ times smaller
scales). It is further important to observe that all the curves for
different $K$ collapse almost perfectly. This means that the energy
transfer due to vortex stretching occurs in a self-similar manner, and 
that the energy is conserved by this cascading process. In other words, 
the viscous effect is negligible in this wavenumber range (i.e. the inertial range). We may verify
the self-similarity in figure \ref{f:scalings}, where we show
$K'_\text{max}/K$ and $K'_\text{min}/K$ as functions of $K$. Here, 
$K'_\text{max}$ and $K'_\text{min}$ denote the wavenumbers which 
attain the maximum and minimum of the energy transfer:
\begin{equation}
 K'_\text{max}(K,t)
 :=
 \mathop{\text{argmax}}_{K'} T(K,K',t)
 \quad\text{and}\quad
 K'_\text{min}(K,t)
 :=
 \mathop\text{argmin}_{K'} T(K,K',t)\:.
\end{equation}
We can see that
$K'_\text{max}\sim K$ and $K'_\text{min}\sim K$ in the wavenumber range
$K\in[0.01\eta^{-1},0.04\eta^{-1}]$.

Thirdly, we look at the dissipation range. Figure \ref{f:T}(c) shows
the energy transfer in a higher wavenumber range:
$K\in[0.031\eta^{-1},0.16\eta^{-1}]$. The behaviors of the energy
transfer are qualitatively different from those in the inertial
range. (i) There is no collapse of the curves because the amount of
energy transfer decreases with wavenumber, since the energy is
dissipated due to viscosity. (ii) the valleys of the energy transfer 
located always at $K'=1.7K$, whereas the wavenumber $K'$ giving the
maximum behaves in a non-trivial manner
 (see figure
\ref{f:scalings}). This implies the disruption of the self-similarity,
which is also reasonable because of the viscous effects.

Fourthly, we look at the far dissipation range. Figure \ref{f:T}(d)
shows the energy transfer in $K\in[0.19\eta^{-1},0.33\eta^{-1}]$. The 
behaviors of the energy transfer in this high wavenumber range is
further different from those in figure \ref{f:T}(c). Though the
Fourier modes acquire the energy form larger scales, almost no energy 
transfers to higher wavenumbers.

\begin{figure}
\begin{center}
\includegraphics[bb=0 0 429 593,width=0.8\textwidth]{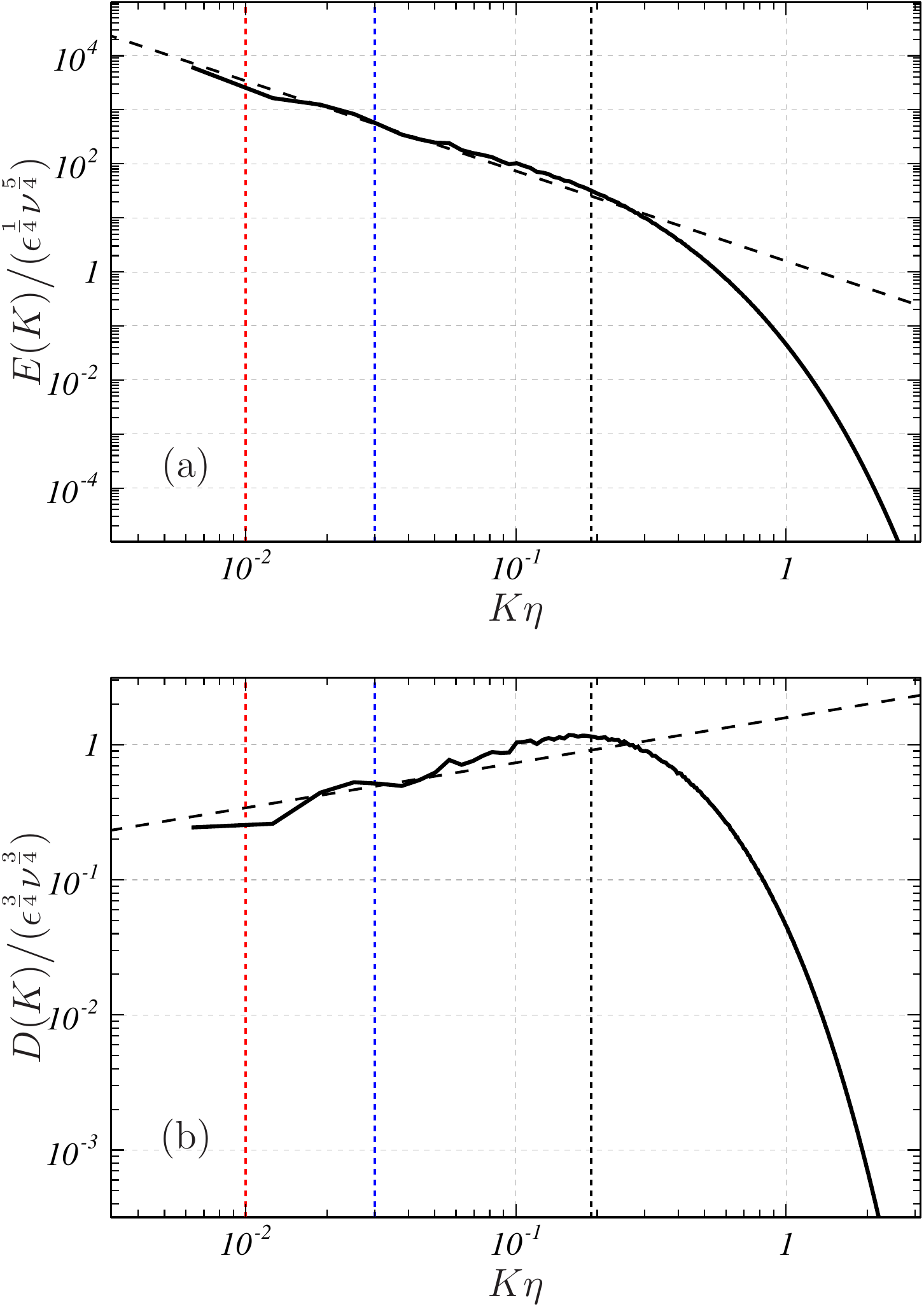}
\end{center}
\caption{(a) Energy and (b) energy dissipation spectrum 
(Results of Run 2-1K; $R_\lambda\approx280$).
Red, blue and black dotted vertical lines indicate
$K=0.01\eta^{-1}$ (the lower end of the inertial range)
$0.03\eta^{-1}$ (the lower end of the dissipation range)
and 
$0.19\eta^{-1}$ (the lower end of the far dissipation range).
Black dashed lines indicate (a) $-5/3$ and (b) $1/3$ power laws.
}
\label{f:spectra}
\end{figure}

Thus, the energy transfer behaves differently in the four wavenumber 
ranges. The bounds of these ranges are located at $K=0.01\eta^{-1}$,
$0.03\eta^{-1}$ and $0.19\eta^{-1}$. These wavenumbers are shown in 
figure \ref{f:spectra}, in which we plot the energy and its
dissipation spectra: 
\begin{equation}\label{usual spectra}
 E(K,t)
 :=
 \oint_{K\le|K'|<K+1}|\widehat{u}(K',t)|^2\:d\Omega
 \quad\text{and}\quad
 D(K,t)
 :=
 \nu K^2 E(K,t)
\end{equation}
The wavenumbers $K=0.01\eta^{-1}$
and $0.03\eta^{-1}$ correspond to the lower and higher ends of the $-5/3$
power law, respectively. We expect that the lower bound normalized by 
$\eta^{-1}$ depends on the 
Reynolds number, whereas the higher bound does not. Furthermore, the
boundary $K=0.19\eta^{-1}$ between the dissipation
and far dissipation ranges is located at the peak of the energy
dissipation spectrum. In summary, we can describe the scale-local energy
transfer once we define it by using the enstrophy production rate
as in \eqref{energy transfer due to vortex stretching}. 

\subsubsection{
Enstrophy production rate in the far dissipation range}\label{far dissipation range}

\begin{figure}
\begin{center}
\includegraphics[bb=0 0 429 293,width=0.8\textwidth]{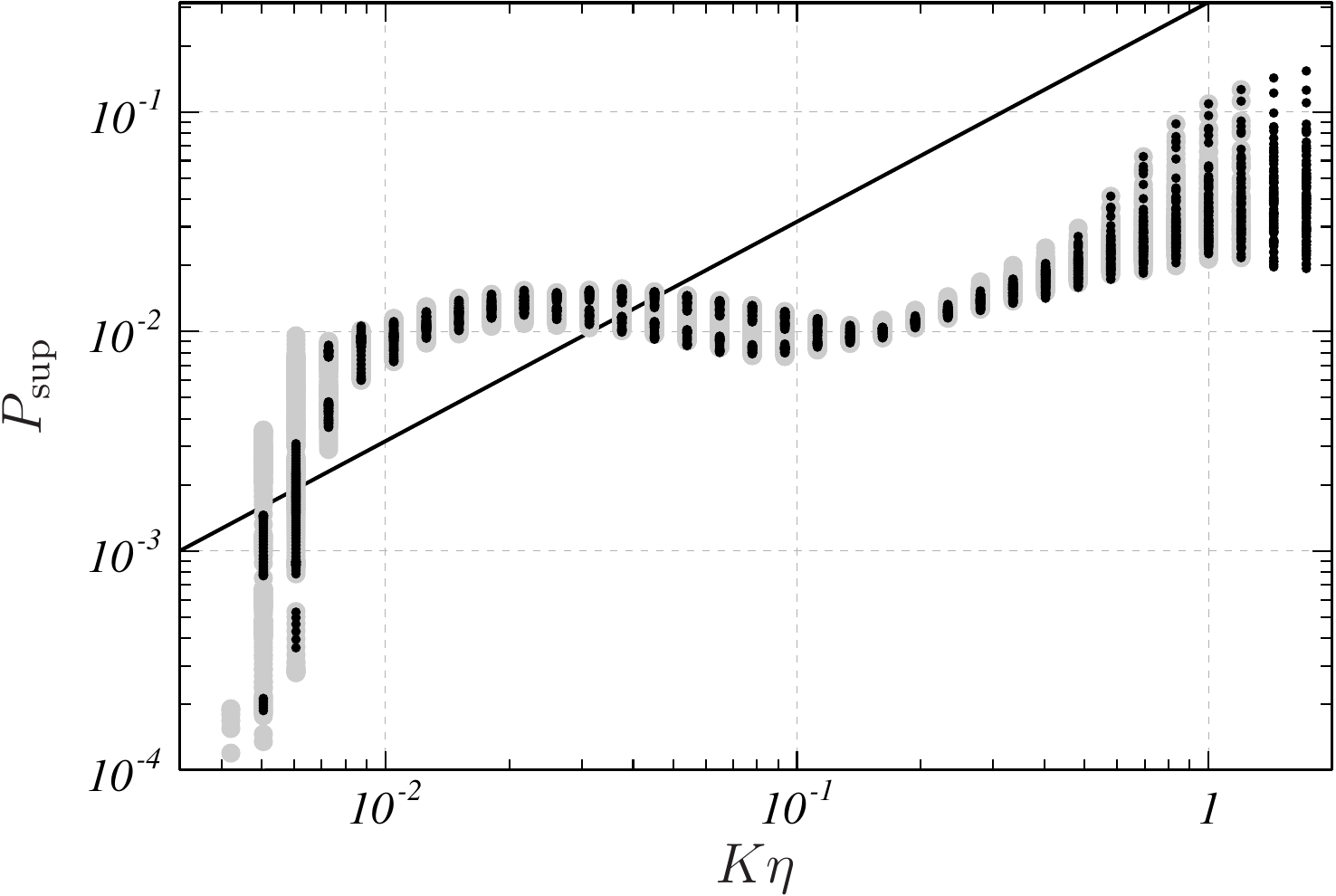}
\end{center}
\caption{
Maximum $P_{\text{sup}}$, defined by (\ref{eq:Psup}), of the normalized
enstrophy production rate. We show the results with two different numerical 
resolutions with a common value of the kinematic viscosity $\nu$;
gray symbols, $N^3=1024^3$, Run 8-1K, $R_\lambda\approx120$;
black symbols, $N^3=2048^3$, Run 8-2K, $R_\lambda\approx130$.
Solid black line indicates $\beta=1$.}
\label{f:sup}
\end{figure}

The results in the previous subsection clarify the usefulness of the 
energy transfer $T(K,K',t)$ defined by the enstrophy production rate
in the description of the energy cascading process. This encourages us
to estimate the maximum enstrophy production rate in the far dissipation
range in order to investigate the possibility of the blowup of solutions to the
Navier-Stokes equation (see Sec. \ref{regularity-criterion}).

For this purpose (i.e.~to investigate the dynamics in the far
dissipation range), we use the DNS results with larger $\nu$ (Runs 8-1K
and 8-2K). Recall that we have conducted DNS with a common value of $\nu$
but with different resolutions $N^3$. 
We plot in figure \ref{f:sup} 
the maximum 
\begin{equation}
\label{eq:Psup}
 P_{\text{sup}}(K,t)
 := 
 \mathop{\text{sup}}_{K'} P(K,K',t)
\end{equation}
of the normalized enstrophy production rate
\begin{equation}\label{normalized enstrophy production rate}
 P(K,K',t)
 :=
 \frac{\displaystyle
 \int_{\mathbb{T}^3}
 \omega(K,x,t)\cdot\nabla u(K',x,t)\cdot\omega(K,x,t)dx
 }
 {\|\omega(K,t)\|_{L^2}^2\:\|\omega(K',t)\|_{L^2}}.
\end{equation}

The different symbols in the figure denote the results with different
resolutions. We may confirm that the results are statistically
independent of the resolution $N^3$, although the instantaneous values
are, of course, dependent on the runs. In particular, we observe that
the temporal fluctuation of $P_\text{sup}$ is quite large in the far
dissipation range ($K>0.1\eta^{-1}$), but the amplitude of the
fluctuations is also independent of the resolution.

The solid straight line shown in this log-log plot indicates 
$\beta=1$ in the main theorem.
The plotted
function seems a concave function of $K$ in the far
dissipation range,
in particular, it seems to satisfy
\begin{equation*}
P_{\text{sup}}(K)\leq CK^\beta
\end{equation*}
for some $\beta<1$ and $C>0$ which is independent of $K$.
This result encourages us to develop the 
mathematics in terms of the energy transfer due to the vortex
stretching. However, the estimation of $P_{\text{sup}}$ in the further
higher dissipation range is numerically challenging because of the 
finite digits of the numerical accuracy, and we must leave this 
for a future numerical study.

\section{Global existence of smooth flow on the NS turbulence.}\label{regularity-criterion}

In this section, 
with the aid of concavity of the normalized enstrophy production rate $P_{\text{sup}}$
 (defined in \eqref{eq:Psup})
in the far dissipation range,  we show that the corresponding solution to the Navier-Stokes equation \eqref{NS} is smooth enough. 
Let us define the inhomogeneous Sobolev  spaces $H^s$ as follows:
\begin{equation*}
H^s(\mathbb{T}^3):=\bigg\{ u=\sum_{K\in\mathbb{Z}^3}\hat u(K)e^{iK\cdot x}\,\bigg| \,\|u\|_{H^s}:=\Big( \sum_{K\in\mathbb{Z}^3}(1+|K|^2)^{s}|\hat u (K)|^2\Big) ^{1/2}<\infty \bigg\},
\end{equation*}
and let us define the band-pass filter (Littlewood-Paley decomposition) as follows:
\begin{equation*}
\omega^k:=\mathcal{F}^{-1}\left[\chi_{[\frac{2^k}{\sqrt{2}},\sqrt{2}2^k)}\mathcal{F}[\omega]\right],
\end{equation*}
where $\chi$ is the characteristic function.
Then, if the vorticity $\omega$ is  mean zero, we can decompose it as follows:
\begin{equation*}
\omega=\sum_{k\in\mathbb{Z}_{\geq -1}}\omega^k
\end{equation*}
and we see
\begin{equation*}
\|\omega\|_{H^s}\approx\left(\sum_{j\in\mathbb{Z}_{\geq -1}}2^{2sj}\|\omega^j\|_{L^2}^2\right)^{1/2}.
\end{equation*}
Now we decompose the vortex stretching term as follows:
\begin{equation}\label{turbulence-assumption}
\begin{split}
\int_{\mathbb{T}^3}(\omega\cdot \nabla)u\cdot \omega 
&=
\sum_{k\in\mathbb{Z}_{\geq -1}}\sum_{m\in\mathbb{Z}_{\geq -1}}\sum_{\ell\in\mathbb{Z}_{\geq -1}}
\int_{\mathbb{T}^3}
(\omega^\ell\cdot\nabla)u^m\cdot\omega^k\\
&=
\sum_{k\in\mathbb{Z}_{\geq -1}}\sum_{m\in\mathbb{Z}_{\geq -1}}\sum_{\ell\in\mathbb{Z}_{\geq -1}}
\widetilde P(2^k,2^m,2^\ell,t)
\|\omega^{\ell}\|_{L^2}\|\omega^m\|_{L^2}\|\omega^k\|_{L^2},\\
\end{split}
\end{equation}
where $\widetilde P$ is defined as follows:
\begin{equation}
\widetilde P(2^k,2^m,2^\ell,t)
 :=
 \frac{\displaystyle
 \int_{\mathbb{T}^3}
 \omega^{\ell}(x,t)\cdot\nabla u^m(x,t)\cdot\omega^k(x,t)dx
 }
 {\|\omega^\ell(t)\|_{L^2}\:\|\omega^m(t)\|_{L^2}\|\omega^k(t)\|_{L^2}}.
\end{equation}
\begin{remark}\label{widetilde P}
By DNS, we have already observed that 
\begin{equation*}
\widetilde P_{\text{sup}}\sim P_{\text{sup}},
\end{equation*}
where
\begin{equation*}
\widetilde P_{\text{sup}}(K,t)=\sup_{K',K''}\widetilde P(K,K',K'',t).
\end{equation*}
Physically, the coincidence of $P_{\text{sup}}$ and $\widetilde P_{\text{sup}}$ is an important issue, thus, as the
sequence of this study, we will provide another research article elsewhere.
\end{remark}

We assume there exist $\beta<1$ and a constant $C>0$ (independent of $k$) such that 
\begin{equation}\label{tilde P}
\sup_{t\in[0,t_*)}\sup_{\ell,m}\widetilde P(2^k,2^m.2^\ell,t)\leq C2^{k\beta}
\end{equation}
as long as the solution $u(t)$ exists in $t\in[0,t_*)$.
By the DNS result in Subsection \ref{far dissipation range},  
 it seems that the NS turbulence satisfies \eqref{tilde P}.
Then we obtain the following regularity criterion. 
\begin{thm}\label{main}
Let $f\in L^\infty(0,\infty ; H^{-1}(\mathbb{T}^3))$ and $u_0\in H^1(\mathbb{T}^3)$ with $
\int_{\mathbb{T}^3} f=\int_{\mathbb{T}^3} u_0=0$ and 
$\nabla\cdot f=\nabla\cdot u_0=0$.
Suppose that $u\in C([0,t_*):H^1(\mathbb{T}^3))$  is a local-in-time strong solution to the three-dimensional Navier-Stokes equation \eqref{NS}.
If
$\beta<1$ for $t\in[0,t_*)$, 
then
$u$ can be extended to the strong solution up to the time $t_{**}$ with $t_{**}>t_*$.
\end{thm}

\begin{remark}\label{smooth-estimate}
By applying  a bootstrapping argument for regularity (see the proof of Theorem 2 in \cite{YY} for example)
and another bootstrapping argument for existence time (see the proof of Theorem 4 in \cite{Y} for example),  
combining with the a-priori bound \eqref{a-priori bound}, we can show that 
the solution exists in an arbitrary time interval, and is smooth enough in both space and time, if the external force $f$ is smooth enough.

\end{remark}

\begin{proof}
The proof is rather elementary.
Our calculation is simply based on \cite[Theorem 3.2]{BMN} and \cite[Section 3.3]{KY}.
They considered global existence of solutions to the incompressible 3D Navier-Stokes equation with the Coriolis force in a periodic domain.
The key point is to construct an a-priori estimate in $H^1$.
First, we estimate \eqref{turbulence-assumption}.
Let us choose $\delta>0$ such that $\beta+\delta<1$.
By Bony's paraproduct formula (see \cite{B} for example), combining H\"older's inequality for sums, we have 
\begin{equation*}
\begin{split}
\int_{\mathbb{T}^3}(\omega\cdot \nabla)u\cdot \omega 
&\leq
C
\sum_{k\in\mathbb{Z}_{\geq -1}}
\sum_{(\ell,m)\in D_1(k)}
2^{\beta k}
\|\omega^{\ell}\|_{L^2}\|\omega^m\|_{L^2}\|\omega^k\|_{L^2}\\
&+
C
\sum_{k\in\mathbb{Z}_{\geq -1}}
\sum_{(\ell,m)\in D_2(k)}
2^{\beta k}
\|\omega^{\ell}\|_{L^2}\|\omega^m\|_{L^2}\|\omega^k\|_{L^2}\\
&+
C
\sum_{k\in\mathbb{Z}_{\geq -1}}
\sum_{(\ell,m)\in D_3(k)}
2^{\beta k}
\|\omega^{\ell}\|_{L^2}\|\omega^m\|_{L^2}\|\omega^k\|_{L^2},\\
\end{split}
\end{equation*}
where (this is well-known Bony's paraproduct formula)
\begin{equation}\label{bony}
\begin{cases}
D_1(k)
&:=\{(\ell,m)\in(\mathbb{Z}_{\geq -1})^2: k-1\leq m\leq k+1,\quad \ell\leq m-1\},\\
D_2(k)
&:=\{(\ell,m)\in(\mathbb{Z}_{\geq -1})^2: k-1\leq \ell\leq k+1,\quad m\leq \ell-1\},\\
D_3(k)
&:=
\{(\ell,m)\in(\mathbb{Z}_{\geq -1})^2:\ell\geq k-1,\quad \ell-1\leq m\leq \ell+1\}.\\
\end{cases}
\end{equation}
Then direct calculations yield
\begin{equation*}
\begin{split}
&\lesssim
\|\omega\|_{L^2}\sum_{k\in\mathbb{Z}_{\geq -1}}2^{\beta k}\|\omega^k\|_{L^2}
\sum_{\ell\leq k}
\|\omega^{\ell}\|_{L^2}\\
&+
\|\omega\|_{L^2}\sum_{k\in\mathbb{Z}_{\geq -1}}2^{\beta k}\|\omega^k\|_{L^2}
\sum_{m\leq k}
\|\omega^m\|_{L^2}\\
&+
\sum_{k\in\mathbb{Z}_{\geq -1}}2^{\beta k}\|\omega^k\|_{L^2}
\sum_{\ell\geq k-1}
\|\omega^{\ell}\|_{L^2}(\|\omega^{\ell-1}\|_{L^2}+\|\omega^\ell\|_{L^2}+\|\omega^{\ell+1}\|_{L^2})\\
&\lesssim
\|\omega\|_{L^2}^2
\left(\sum_{k\in\mathbb{Z}_{\geq -1}}2^{2(\beta+\delta)k}\|\omega^k\|_{L^2}^2\right)^{1/2}
\left(\sum_{k\in\mathbb{Z}_{\geq -1}}2^{-2\delta k}k\right)^{1/2}\\
&+
\|\omega\|_{L^2}^2
\left(\sum_{k\in\mathbb{Z}_{\geq -1}}2^{2(\beta+\delta)k}\|\omega^k\|_{L^2}^2\right)^{1/2}
\left(\sum_{k\in\mathbb{Z}_{\geq -1}}2^{-2\delta k}\right)^{1/2}\\
&\lesssim_\delta
\|\omega\|_{H^{\beta+\delta}}\|\omega\|_{L^2}^2.\\
\end{split}
\end{equation*}
Here we used the fact that 
\begin{equation*}
\sum_{\ell\leq k}a_\ell\leq \left(\sum_{\ell\leq k}a_\ell^2\right)^{1/2}\left(\sum_{\ell\leq k}1\right)^{1/2}\quad\text{for}\quad a_\ell\geq 0.
\end{equation*}
 We just  proceed the $H^1$ energy estimate as (recall $\|\omega\|_{L^2}\approx\|u\|_{H^1}$ with $\int_{\mathbb{T}^3}u=0$) 
\eqq{\frac{d}{dt}\norm{u(t)}{H^1}^2+2\nu \norm{u(t)}{H^{2}}^2\lesssim \norm{u(t)}{H^1}^2
\|u(t)\|_{H^{1+\beta+\delta}}+\|f\|_{H^1}\|u\|_{H^1}.}

By interpolation and Young's inequality (note $\beta+\delta\leq1$), 
\begin{equation*}
\begin{split}
&\norm{u}{H^1}^2\|u\|_{H^{1+\beta+\delta}}\lesssim
\norm{u}{H^1}^2\|u\|_{H^{2}}
\lesssim \nu^{-1} \norm{u}{H^{1}}^4+\nu\norm{u}{H^2}^2,\\
&
\|f\|_{H^1}\|u\|_{H^1}\lesssim \|f\|_{H^1}^2+\|u\|_{H^1}^2,
\end{split}
\end{equation*}
and hence,
\eqq{\frac{d}{dt}\norm{u(t)}{H^1}^2+\nu \norm{u(t)}{H^{2}}^2
\le (1+\nu ^{-1}\norm{u(t)}{H^1}^{2})\|u(t)\|_{H^1}^2+\|f\|_{H^1}^2, \qquad t>0.}
On the other hand, we immediately  have the energy inequality:
\begin{equation*}
\frac{d}{dt}\norm{u(t)}{L^2}^2+2\nu \norm{u(t)}{H^{1}}^2\lesssim \|f(t)\|_{H^{-1}}\|u(t)\|_{H^{1}}.
\end{equation*}
Thus, by the absorbing argument, we have 
\begin{equation*}
\norm{u(t)}{L^2}^2+\nu \int_0^t\norm{u(t)}{H^{1}}^2\lesssim \nu^{-1}\int_0^t\|f(t)\|_{H^{-1}}^2+\|u(0)\|_{L^2}^2.
\end{equation*}
By the Gronwall inequality, we then have 
\begin{equation}\label{a-priori bound}
\begin{split}
&\|u(t)\|_{H^1}^2\lesssim\|u_0\|_{H^1}^2e^{\int_0^t(1+\nu^{-1}\|u(s)\|_{H^1}^2)ds}
+\int_0^te^{\int_\tau^t(1+\nu^{-1}\|u(s)\|_{H^1}^2)ds}\|f(\tau)\|_{H^1}^2d\tau\\
&
\lesssim
\|u_0\|_{H^1}^2\exp\left(t+\nu^{-2}\int_0^t\|f(s)\|_{H^{-1}}^2ds+\nu^{-1}\|u_0\|_{L^2}^2\right)\\
&
+\exp\left(t+\nu^{-2}\int_0^t\|f(s)\|_{H^{-1}}^2ds+\nu^{-1}\|u_0\|_{L^2}^2\right)
\int_0^t\|f(\tau)\|_{H^1}^2d\tau<\infty
\end{split}
\end{equation}
for  $t\in[0,t_*)$.
By this a-priori estimate combining local existence result (see \cite[Theorem 1.2]{KY} for example), we can immediately prove the main theorem.

\end{proof}

\section{Reformulation of the Kolmogorov-Richardson energy cascade}\label{sophisticated}

In this section, with the aid of the DNS result in Sec. \ref{s:DNS},  we describe
the Kolmogorov-Richardson energy cascade in terms of vortex stretching.
In what follows (including the next section),  we employ $\mathbb{R}^3$ which is as an approximation domain of $\mathbb{T}^3$. 
Also we simply write $\int f:=\int_{\mathbb{R}^3}f(x)dx$.
Let $u$ be a smooth solution to the 3D-Navier-Stokes equation \eqref{NS} (but replace $\mathbb{T}^3$ to $\mathbb{R}^3$), and $\omega$ be the corresponding vorticity.
Let
\begin{equation}\label{LP-decomposition}
\begin{split}
\bar\omega_K(t,x)
&:=\mathcal{F}^{-1}_\xi[\chi_{A_K}(\xi)\hat \omega(t, \xi)](x),
\end{split}
\end{equation}
where (coarse-graining in an annulus, Littlewood-Paley decomposition)
\begin{equation*}
A_K:=\{\xi\in\mathbb{R}^3:K/\sqrt \alpha\leq |\xi|\leq \sqrt \alpha K\}
\end{equation*}
and $\alpha>1$ is a prescribed constant expressing the ratio of 
adjacent scales (see 
Figure \ref{f:T}).
In what follows we take wavenumber $K$ 
 from  $\alpha^{\mathbb{Z}}$. Also, in the following arguments, we assume that the spatially integrated value of any quantity is independent of time because the size of the domain is much larger than the correlation length (the integral length) of the flow, that is, 
\begin{equation}\label{statistically stationary}
\frac{d}{dt}\|\bar \omega_K(t)\|_{L^2}^2=0\quad\text{for any}\quad K\quad\text{and}\quad t\geq 0.
\end{equation}
Our specific purpose in this section is to approximate the following vortex stretching term:
\begin{equation*}
\int(\omega\cdot\nabla)u\cdot \bar \omega_K.
\end{equation*}
Note that $\omega=\sum_{K\in \alpha^{\mathbb{Z}}}\bar\omega_K$ and $u=\sum_{K\in \alpha^{\mathbb{Z}}}\bar u_K$.
First, we propose a new hypothesis.
For the original version of Kolmogorov hypothesis, see \cite{K41}, for a recasted version, see \cite[Section 6]{F}.

\vspace{0.3cm}

\noindent
{\bf Outline of our hypothesis.}\quad
The crucial point of our hypothesis is that there exists a universal vortex stretching/compressing mechanism, independent of the energy input rate $\epsilon$ (which will appear later).
More precisely, a pair of
tubular vortices $W_{j,K}$ 
are stretching  
 several tubular vortices in the adjacent smaller scale, or,
several tubular vortices are compressing a pair of tubular vortices in the adjacent larger scale.
Also, supports of these tubular vortices (with compact supports) are disjoint. We conjecture that such vortex stretching/compressing in adjacent two scales are the dominant event, so,
in our cascade picture, we exclude multiscale events. 
Note that, by Remark \ref{three wave interaction}, due to the uncertainty principle for the Fourier transform, we expect that 
the three-wave interaction of this adjacent scale event is rather nonlocal (c.f. \cite{OK}).

\vspace{0.3cm}

\noindent
{\bf Balance of vortex stretching/compressing.}\quad 
Vortex stretching and compressing
are balanced if \eqref{statistically stationary} holds. See also \eqref{key-estimate}.

\vspace{0.3cm}

\noindent
{\bf A pair of tubular vortices.}\ 
Let $W_{j,K}\in C^\infty_c(\mathbb{R}^3)$ represents  a pair of ``normalized" tubular vortices, where $j$ and $K$ are important indexes, will appear later.
First  we reasonably assume 
\begin{equation}\label{emptyset}
\text{supp}\, W_{j,K}\cap\text{supp}\, W_{j',K}=\emptyset\quad\text{for}\quad j\not=j'.
\end{equation}
Next we define $\widetilde W_{j,K}(x)$ as follows: 
\begin{equation*}
\widetilde W_{j,K}(x):=\mathcal{F}^{-1}\left[\chi\mathcal{F}[W_{j,K}]\right](x),
\end{equation*}
 where
$\chi$ is a characteristic function such that 
\begin{equation*}
\chi(\xi)=
\begin{cases}
\displaystyle 1\quad \frac{1}{\sqrt \alpha}\leq |\xi|\leq \sqrt \alpha,\\
0\quad \text{otherwise}.
\end{cases}
\end{equation*}
Note that $\text{supp}\,\mathcal{F}[\widetilde W_{j,K}
]\subset A_1$
and  $\widetilde W_{j,K}$ is decaying (not compactly supported) in the physical space, due to the uncertainty principle.
We  assume that
\begin{equation*}
\displaystyle\int_{\mathbb{R}^3} |W_{j,K}(x)
|^2dx=\gamma
\end{equation*}
for some prescribed (non-dimensional) constant $\gamma$
 which is independent of 
 all parameters. 
\begin{remark}
In the real turbulence, this $\gamma$ is rather probability distribution.
\end{remark}
By Parseval's identity, we see that 
\begin{equation*}
\displaystyle\int_{\mathbb{R}^3} |\widetilde W_{j,K}(x)
|^2dx=\int_{\mathbb{R}^3} |W_{j,K}(x)|^2dx+\int_{\mathbb{R}^3}|\widetilde W_{j,K}(x)-W_{j,K}(x)|^2dx.
\end{equation*}
Also assume these $\widetilde W_{j,K}$ and $W_{j,K}$ are close to each other, that is, 
\begin{equation}\label{smallness}
\frac{\displaystyle\int_{\mathbb{R}^3} |W_{j,K}(x)-\widetilde W_{j,K}(x)|^2dx}{\displaystyle\int_{\mathbb{R}^3}|W_{j,K}(x)|^2dx}\leq \delta< 1
\end{equation}
for some small $\delta>0$.
In Appendix, we give a typical example of tubular vortices $W_{j,K}$, and we estimate $\delta$ by using it.


\vspace{0.3cm}

\noindent
{\bf Decomposition of vorticity field.}\
In the inertial range, 
 we employ the following hypothesis.
For any fixed $t$, assume that
 $\bar\omega_K$ is mainly expressed by $\{W_{j,K}\}_{j}$, and is independent of $\nu$ and $\delta$,  as follows:
\begin{equation}\label{scaling assumption}
\begin{split}
\bar\omega_K(t,x)
&
=
g K^{-H+1}\left(\sum_{j=1}^{cK^D
}
W_{j,K}(Kx)
\right),\\
\end{split}
\end{equation}
where $c, g, H, D\in\mathbb{R}_{\geq 0}$. 
Rigorously, there must exist a small perturbation (in order to recover divergence-free
and compact support in the Fourier space) depending on $\nu$ and $\delta$, but we neglect it.
$H$ and $D$ express the H\"older exponent
and the fractal dimension respectively.
Rigorously, we need to require $cK^D$ to be integer. In this case we slightly approximate 
$c$ and $D$. 
This  $D$ is ideally determined by physical experiments. 
For example, in \cite{TGOY}, we have investigated this exponent by numerical
computations. In particular, we identified the tubular vortices in each scales by the low-pressure method, and we figured out that the hierarchy of stronger tubular vortices is indeed intermittent with the dimension $D$ smaller than 3.
The above assumptions are supported by \cite{TGOY} in some extent.

\begin{remark}(The specific aim of this paper.)\quad
The vorticity version of the Littlewood-Paley spectra $E_{LP}(K)$ is given by
\begin{equation*}
E_{LP}(K):=K^{-3}\|\bar\omega_K\|_{L^2}^2
=c\gamma g^{2}K^{-\frac{5}{3}}K^{-\frac{1}{3}(3-D)}
K^{\frac{2}{3}(-3H-2+D)}
\end{equation*}
(c.f. Sec. 2 in \cite{Constantin} for the definition of the Littlewood-Paley spectra).
Thus, our specific aim is to figure out the constants $g$ and $H$.
Note that this $E_{LP}$ dimensionally coincides with another $E$ in \eqref{usual spectra}.
\end{remark}


\vspace{0.3cm}
\noindent
{\bf The external force.}\quad
For fixed $c>0$, we assume that, for any energy input rate 
$\epsilon$ and wavenumber $K_f$, we can construct an external force $f$ such that 
$\text{supp}\, \hat f\subset A_{K_f}$ and 
\begin{equation*}
\frac{1}{K_f^2}\
\int\overline{(\nabla\times f)}_{K_f}\cdot \omega
=\epsilon, 
\end{equation*} 
where $\omega$ is a solution to the Navier-Stokes (vorticity) equation.


\vspace{0.3cm}

\noindent
{\bf Hierarchy of tubular vortices.}\quad 
Let $K_{\pm}=\alpha^{\mp 1} K$
 and
let us assume that
a pair of
tubular vortices $W_{J,K_+}$
 stretches (through Biot-Savart law)
$W_{j,K}$ ($j\in^{\exists}\!\Omega_{J,K}^{+}\subset\{1,2,\cdots, cK^D\}$) and assume disjointness
 $\Omega_{J,K}^{+}\cap\Omega_{J',K}^{+}=\emptyset$ ($J\not=J'$).
At the same time, this $W_{j,K}$ compresses $W_{J,K_+}$. If $W_{J,K_+}$ does not stretch anything, then we regard $\Omega_{J,K}^{+}$ as an empty set. 
In the forcing scale, the external force $\nabla\times f$ stretches $W_{j,K_f}$
$(j\in ^\exists\!\Omega_{K_f}^+\subset\{1,2,\cdots,c(K_f)^D\})$.
Let 
\begin{equation*}
\Omega_K^{+}=\bigcup_{J=1}^{cK_+^D}\Omega_{J,K}^{+}.
\end{equation*}
 To the contrary, assume that, 
for any $j \in (\Omega_{K}^{+})^c:=\{1,2,\cdots, cK^D\}\setminus\Omega_K^{+}$,
$W_{j,K}$   is compressed by  $W_{j',K_-}$ ($j'\in^{\exists}\!\Omega_{j,K}^{-}\subset\mathbb{N}$) through Biot-Savart law. 
At the same time, $W_{j',K_-}$ is stretched by this $W_{j,K}$.
Note that
$\Omega_{J,K}^+=\Omega_{J,K_+}^-$ and  $|\Omega_{K}^{+}|+|(\Omega_K^{+})^c|=K^D$.
The construction of this scenario is inspired by Movie S9 in ``Supplementary materials" in \cite{MOMPBR}.
In this movie, we can observe forming an ordered array of counter-rotating secondary vortex filaments perpendicular to the primary cores. We can also  observe secondary filaments interacting to form a new generation of perpendicular tertiary vortex filaments.

\vspace{0.3cm}
\noindent
{\bf Clarification of $\alpha$ and $|\Omega_{j,K}^+|$.}\quad
We crucially assume $\alpha$ and $|\Omega_{j,K}^{+}|(=|\Omega_{j,K}^-|)$ are  non-dimensional prescribed constants (ideally determined by physical experiments, see also Figure \ref{f:T}), that is, independent of $K$ and $j$.
Rigorously this $|\Omega_{j,K}^+|$ is integer, but here we are implicitly 
spatially averaging it,
 to be a positive real number.
We conjecture that this $|\Omega_{j,K}^{+}|$ is determined by 
an unstable mode (which is independent of any type of disturbance) induced by 
the elliptical instability 
(see Subsection 2.2 in \cite{K}, see also \cite{MOMPBR}).
More precisely, the elliptical instability originates from the parametric excitation
of Kelvin modes in the vortex cores in each tubular vortices
$W_{j,K}$.

\vspace{0.3cm}
\noindent
{\bf  Ratios of  the number 
of tubular vortices: being stretched and being compressed.}\quad
Let  $\tau_+(K)$ and $\tau_-(K)$ be ratios of  the number 
of tubular vortices, being stretched and being compressed, respectively.
Since in the statistically steady state, these ratios could be estimated by the ratio of 
turnover time in adjacent scales, it follows that
\begin{equation*}
\text{turnover time}\sim \frac{\text{length}}{\text{velocity}}\sim g^{-1} K^{H-1}.
\end{equation*}
Thus we can define $\tau_{\pm}$ (satisfying $\tau_++\tau_-=1$) as follows:
\begin{equation*}
\begin{split}
&\tau_+(K):=\frac{K^{H-1}}{K^{H-1}+K_-^{H-1}}=\frac{1}{1+\alpha^{H-1}},\\
&\tau_-(K):=\frac{K^{H-1}_-}{K^{H-1}+K_-^{H-1}}
=\frac{1}{\alpha^{-H+1}+1}.
\end{split}
\end{equation*}
Consequently $\tau_{\pm}$ are independent of $K$.

\vspace{0.3cm}
\noindent
{\bf Vortex stretching/compressing relation vs spatial and statistical independence.}\quad
For tubular vortices $W_{j_1,K_1}$ and $W_{j_2,K_2}$, let us define vortex stretching/compressing relations as follows:
\begin{itemize}
\item Vortex stretching relation:
\begin{equation*}
\begin{split}
V^+:=\{(j_1,K_1,j_2,K_2): j_1\in\Omega^+_{j_2,K_1}\  \text{and}\  K_2=\alpha^{-1} K_1\},
\end{split}
\end{equation*}
\item Vortex compressing relation:
\begin{equation*}
\begin{split}
V^-:=\{(j_1,K_1,j_2,K_2): j_2\in\Omega^-_{j_1,K_1}\ \text{and}\  K_2=\alpha K_1\}.
\end{split}
\end{equation*}
\end{itemize}
\begin{definition}\label{disjointness}
(Vortex stretching/compressing relation: disjointness)\quad
We assume the following: If $W_{j_1, K_1}$ and $W_{j_2,K_2}$ 
have vortex stretching/compressing relation, that is, if 
$(j_1,K_1,j_2,K_2)\in V^+\cup V^-$,
then
\begin{equation*}
\text{supp}\,  W_{j_1,K_1}\cap\text{supp}\, W_{j_2,K_2}=\emptyset.
\end{equation*}
\end{definition}
\begin{definition}\label{statistically-independent} (Spatial and statistical independence.)\quad
Let $U_{j,K}:=-\nabla\Delta^{-1}\times W_{j,K}$.
Then we significantly assume the following:
\begin{equation*}
\begin{split}
&\sum_{(j_1,K_1,j_2,K_2)\in (V^+)^c\cap (V^-)^c
} \int\left(
W_{j_1,K_1}
\cdot\nabla\right)
U_{j_2,K_2}
\cdot 
W_{j^*,K^*}
=0\\
\end{split}
\end{equation*}
and
\begin{equation*}\label{vortex stretching relation case}
\begin{split}
&\sum_{
\stackrel{(j_1,K_1,j_2,K_2)\in V^+\cup V^-,}{(j_1,K_1)\not=(j^*,K^*)\ \text{and}\ (j_2,K_2)\not=(j^*,K^*)}}
 \int\left(
W_{j_1,K_1}
\cdot\nabla\right)
U_{j_2,K_2}
\cdot 
W_{j^*,K^*}
=0.
\end{split}
\end{equation*}
for any fixed $j^*$ and $K^*$.
\end{definition}
By Definitions \ref{disjointness} and \ref{statistically-independent}, we have that   
\begin{equation*}
\begin{split}
\int (\omega\cdot\nabla)u\cdot\bar\omega_K
&=
\int(\bar\omega_K\cdot\nabla) \bar u_{K_+}\cdot \bar\omega_K
+
\int(\bar \omega_K\cdot\nabla) \bar u_{K_-}\cdot \bar\omega_K.
\end{split}
\end{equation*}

\vspace{0.3cm}
\noindent
{\bf Normalization of vortex stretching/compressing.}\\
Let $\omega^*_K(x):=\bar\omega_K(x)/(gK^{-H+1})$
and let 
$\omega_K^{**}(x):=\omega_K^*(K^{-1}x)$.
Then we see that
\begin{equation*}
\begin{split}
&\int(\bar\omega_K\cdot\nabla) \bar u_{K_+}\cdot \bar\omega_K
+
\int(\bar\omega_K\cdot\nabla) \bar u_{K_-}\cdot \bar\omega_K\\
=&
g^{3}K^{-3(H-1)}\left(\alpha_+^{-H+1}\int(\omega^*_K\cdot\nabla) u^*_{K_+}\cdot \omega^*_K
+\alpha_-^{-H+1}
\int(\omega^*_K\cdot\nabla)\cdot u^*_{K_-}\cdot \omega^*_K
\right)\\
=&
g^{3}K^{-3(H-1)}K^{-3}\left(\alpha_+^{-H+1}\int(\omega^{**}_K\cdot\nabla) u^{**}_{K_+}\cdot \omega^{**}_K
+\alpha_-^{-H+1}
\int(\omega^{**}_K\cdot\nabla)\cdot u^{**}_{K_-}\cdot \omega^{**}_K
\right).
\end{split}
\end{equation*}
Note that these $\omega^{**}_K$, $u_{K_+}^{**}$ and $u_{K_-}^{**}$ are precisely expressed as 
\begin{equation*}
\begin{split}
\omega^{**}_K(t,x)
&=\sum_{j=1}^{cK^D}
W_{j,K}(x)\\
u_{K_+}^{**}(t,x)
&=\sum_{j=1}^{cK_+^D}U_{j,K_+}(\alpha^{-1}x)=:\sum U_{j,K_+},\\
u_{K_-}^{**}(t,x)
&=\sum_{j=1}^{cK_-^D}U_{j,K_-}(\alpha x)=:\sum U_{j,K_-},\\
\end{split}
\end{equation*}
for each fixed $t$.
By \eqref{emptyset},
then we see that
\begin{equation*}
\begin{split}
&\int(\omega^{**}_K\cdot\nabla) u^{**}_{K_+}\cdot \omega^{**}_K
=
\sum_J\sum_{j\in\Omega_{J,K}^+}\int(W_{j,K}
\cdot\nabla)
U_{J,K_+}
\cdot 
W_{j,K}
,\\
&\int(\omega^{**}_K\cdot\nabla) u^{**}_{K_-}\cdot \omega^{**}_K
=
\sum_{j\in(\Omega_{K}^+)^c}
\sum_{j'\in\Omega_{j,K}^-}\int(W_{j,K}
\cdot\nabla)
U_{j',K_-}
\cdot 
W_{j,K}
,
\end{split}
\end{equation*}
where 
\begin{equation*}
\nabla U_{J,K}:=-\nabla(\nabla\Delta^{-1}\times W_{J,K}).
\end{equation*}
Let $S_{{\pm}}$ and $P_{{\pm}}$ be the strain and rotation tensors of $\nabla u^{**}_{K_{\pm}}$ respectively.
Then we can rewrite the following vortex stretching/compressing terms:
\begin{equation*}
\int(\omega^{**}_{K}\cdot \nabla)u^{**}_{K_{\pm}}\cdot \omega^{**}_{K} =\int(\omega_{K}^{**})^T(P_{{\pm}}+S_{{\pm}})\cdot\omega^{**}_{K}=\int (\omega_{K}^{**})^TS_{\pm}\cdot\omega^{**}_{K},
\end{equation*}
where $(\omega_{K}^{**})^T$ is the row vector of $\omega^{**}_{K}$.
By the direct calculation, the rotation tensor part disappears.
Let $S_{J,K_{\pm}}$ be the rate-of-strain tensor of $\nabla U_{J,K_{\pm}}$.

\vspace{0.3cm}

\noindent
{\bf Clarification of a universality of vortex stretching/compressing.}\quad 
We now clarify the universal vortex stretching/compressing hypothesis.
In the vortex stretching case, we assume
\begin{equation}\label{vortex-stretching}
\Lambda_+:=\int(W_{j,K})^TS_{J,K_+}\cdot W_{j,K}
>0
\quad\text{for}\quad j\in \Omega_{J,K}^+, 
\end{equation}
where $\Lambda_+$ is the absolute constant, in particular, independent of $\epsilon$.
On the other hand, in the vortex compressing case, we assume
\begin{equation}\label{vortex-compressing}
-\Lambda_-:=\int(W_{j,K})^T\sum_{j'\in\Omega_{j,K}^-}S_{j',K_-}\cdot W_{j,K}
<0\quad\text{for}\quad j\in(\Omega_{K}^+)^c,
\end{equation}
where $\Lambda_->0$ is the absolute constant, in particular, independent of $\epsilon$.

\begin{remark}\label{three wave interaction}
Due to the uncertainty principle, vortex stretching in the adjacent scale 
\begin{equation*}
\mathcal{F}\left[\sum_{j\in\Omega^+_{J,K}}(W_{j,K})^TS_{J,K_{+}}\cdot
 W_{j,K}\right]
\end{equation*}
has non-compact support in the Fourier space (by the same reason, the vortex compressing  also has non-compact support).
More precisely we expect that the following three wave interaction is nonzero:
\begin{equation*}
\mathcal{F}\left[\sum_{j\in\Omega^+_{J,K}}\left(\mathcal{F}^{-1}\left[\mathcal{F}[W_{j,K}]\chi_{A_{K_1}}\right]\right)^T\mathcal{F}^{-1}\left[\mathcal{F}[S_{J,K_{+}}]\chi_{A_{K_2}}\right]\cdot
 \mathcal{F}^{-1}\left[\mathcal{F}[W_{j,K}]\chi_{A_{K_3}}\right]\right]
\end{equation*}
for the following nonlocal three wave combinations (c.f. \eqref{bony}):
\begin{equation*}
\begin{cases}
K_1\approx K_3,\quad K_2\ll K_1,\\
K_2\approx K_3,\quad K_1\ll K_2,\\
K_1\approx K_2,\quad K_3\ll K_1.
\end{cases}
\end{equation*}
\end{remark}

Summarizing the above arguments, we have 
\begin{equation*}
\begin{split}
 \int (\omega_K^{**})^T S_{K_{\pm}}\cdot\omega^{**}_K
=\pm c
 K^D\Lambda_{\pm}\tau_{\pm}\alpha^{\mp(-H+1)}.
\end{split}
\end{equation*}
\begin{remark}
We can apply the same calculation to the convection term, and 
clearly it becomes zero:
\begin{equation*}
\begin{split}
&\int(u\cdot\nabla)\omega\cdot\bar\omega_K=
\int(u\cdot\nabla)\bar\omega_K\cdot\bar\omega_K
= 0\\
\end{split}
\end{equation*}
due to skew-symmetry.
\end{remark}
We now summarize the important definitions of ``coherent vortices" as the following:

\vspace{0.3cm}
\noindent 
{\bf Coherent vortices:}\quad
Let us define the approximate parameters:
\begin{equation*}
\begin{split}
C_{+}&:=c
\Lambda_+\tau_+\alpha^{-(-H+1)},\\
C_{-}&:=c
\Lambda_-\tau_-\alpha^{-H+1}.\\
\end{split}
\end{equation*}
Note that the constants $c$, 
 $\Lambda_{\pm}$ and $\alpha$ are independent of $\epsilon$.

\begin{itemize}
\item
Inertial range (c.f. enstrophy production rate \eqref{energy transfer due to vortex stretching})
\end{itemize}
\begin{equation*}
\begin{split}
\frac{1}{K^2}\int(\omega\cdot\nabla)u\cdot\bar\omega_K
=
 g^{3}(C_{+}-C_-)K^{-3H-2+D}
\end{split}
\end{equation*}
for $K_{F}< K< ^\exists\!\!K_{dis}$.
Also, from the physical observation \ref{physical observation}, we mathematically  assume $K_{dis}\to \infty$ as $\nu\to 0$.
\begin{itemize}
\item
Driving force scale (beginning of the coherent vorticity)
\end{itemize}
In the forcing scale, we assume that the effect of stretching is negligible:
 \begin{equation}\label{negligible}
\int(\bar\omega_{K_{f}}\cdot\nabla)\bar u_{(K_{f}/\alpha)}\bar\omega_{K_{f}}= 0.
\end{equation}
In other words,  initially, driving force $f$ 
creates the largest scale vortices. 
Thus
 \begin{equation*}
\frac{1}{K_f^2}\int(\omega\cdot\nabla)u\cdot\bar\omega_{K_f}
=-g^{3}C_-K_f^{-3H-2+D}.
\end{equation*}

\section{Derivation of Kolmogorov's $-5/3$ law.}\label{derivation of K41}

In this section, we derive
Kolmogorov's $-5/3$ law.
The main theorem in this section is the following.
\begin{theorem}
Assume \eqref{statistically stationary} and 
``coherent vortices".
Taking $\nu\to 0$, then we have Kolmogorov's $-5/3$ law:
\begin{equation}\label{K41}
\begin{split}
E_{LP}(K)
:=K^{-3}
\|\bar \omega_K\|_{L^2}^2
=
c\gamma\epsilon^{2/3}K^{-5/3}
 K^{-\frac{1}{3}(3-D)}
\end{split}
\end{equation}
for $K\geq K_f$. 
\end{theorem}



\vspace{0.4cm}
\noindent
{\bf Proof.}\quad
We multiply by $\bar \omega_K$ to \eqref{NS}, integrate on both sides, divided by $K^2$,
  to obtain 
\begin{equation}\label{modern-energy-balance-0}
\begin{split}
&
\frac{1}{2K^2}\frac{d}{dt}\int_{\mathbb{R}^3}|\bar \omega_K|^2 
+
\frac{\nu}{K^2} \int_{\mathbb{R}^3} |\nabla \bar \omega_K|^2
\\
& +
\frac{1}{K^2}\int_{\mathbb{R}^3}
(u\cdot \nabla)\omega
\cdot\bar\omega_K
-
\frac{1}{K^2}\int_{\mathbb{R}^3}
(\omega\cdot\nabla)u
\cdot\bar\omega_K
\\
&=
\frac{1}{K^2}\int_{\mathbb{R}^3} 
(\nabla\times f)
\cdot \bar \omega_K
\end{split}
\end{equation}
for $K\geq K_f$. 
By the assumption: coherent vortices and \eqref{statistically stationary},
we have 
\begin{equation*}
\begin{split}
\frac{\nu}{K^2} \int_{\mathbb{R}^3} |\nabla \bar \omega_K(\cdot,x)|^2dx
- g^{3}(C_{+}-C_-)K^{-3H-2+D}
=0
\end{split}
\end{equation*}
for $K_f<K<K_{dis}$ and 
\begin{equation*}
\begin{split}
\frac{\nu}{K_f^2} \int_{\mathbb{R}^3} |\nabla \bar \omega_{K_f}(\cdot,x)|^2dx
+ g^{3}C_-K_f^{-3H-2+D}
=\epsilon.
\end{split}
\end{equation*}
On the other hand,
\begin{equation*}
\frac{\nu}{K^2}\|\nabla \bar \omega_K(\cdot)\|_{L^2}^2
\approx 
\nu\|\bar\omega_K(\cdot)\|_{L^2}^2
=
\nu c\gamma g^{2}K^{-2H-1+D}. 
\end{equation*}
Thus
\begin{equation*}
|C_+-C_-|\lesssim \nu c\gamma g^{-1}K^{H+1}
\end{equation*}
for $K_f<K<K_{dis}$ and 
\begin{equation*}
|\epsilon-C_-g^{3}K_f^{-3H-2+D}|<\nu c\gamma g^{2}K_f^{-2H-1+D}.
\end{equation*}
Since $C_{\pm}$, $M$ and $H$ are independent of $\nu$ and $K$,
 we have 
\begin{equation}\label{key-estimate}
C_+=C_-\quad\text{and}\quad
\epsilon=C_-g^{3}K_f^{-3H-2+D}
\end{equation}
for taking $\nu$ to zero. 
Then the rest argument is rather obvious.
First, assume $H\not=(-2+D)/3$, 
then it contradicts \eqref{key-estimate}
if we take  $K_{f}$ sufficiently large.  
Thus, we have $H=(-2+D)/3$.
Consequently, $H$ (and also $\tau_{\pm}$) is independent of $\epsilon$.
Second, assume $g\not=\epsilon^{1/3}$,
then it contradicts  \eqref{key-estimate}
if we take  $\epsilon$ sufficiently large. Thus, $g=\epsilon^{1/3}$. Consequently, $C_-=C_+=1$ and then $\Lambda_{\pm}$ are also determined.
Therefore the Littlewood-Paley spectra $E_{LP}(K)$ is  given by
\begin{equation*}
E_{LP}(K):=K^{-3}
\|\bar\omega_K\|_{L^2}^2
=
c\gamma\epsilon^{2/3}K^{-5/3} K^{-\frac{1}{3}(3-D)}.
\end{equation*}
This is Kolmogorov's $-5/3$ power law.


\section{Conclusion}

We investigated the hierarchy of coherent vortices observed in the DNS of forced turbulence in a periodic domain. First, we proved that if the normalized enstrophy production rate $P_{\text{sup}}$ is concave in the far dissipation range (more precisely, $\beta<1$), the corresponding solution to the Navier-Stokes equation is smooth enough. Since the DNS results shown in figure 4 seem to support the new regularity criterion that $P_{\text{sup}}$ is concave in the far dissipation range, we may conclude that the turbulence driven by the deterministic force in a periodic domain is smooth enough. Second, we construct the hierarchy of coherent vortices which is statistically self-similar in the inertial range so that we can estimate the energy transfer rate in adjacent scales due to the vortex stretching/compressing process. The crucial assumption, which is clearly supported by DNS results (figure 2), of this construction is that the process is self-similar and local in scale. Then, under
 this assumption, we may derive Kolmogorov's $-5/3$ power law of the energy spectrum of statistically stationary turbulence. We emphasize that our derivation does not require Kolmogorov's hypotheses, but instead it requires a universal mechanism of the vortex stretching and compressing. Though the present study lacks the discussion on the origin of this universality, the recent study by McKeown {\it et al.} \cite{MOMPBR} may provide with it; they investigated how the elliptical instability led to the persistence of the turbulent energy cascade through the local interactions of vortices over a hierarchy of scales.

\section{Appendix: a typical example of antiparallel tubular vortices}
In this section we give a typical example of a pair of antiparallel tubular vortices,
and
estimate $\delta$ in \eqref{smallness}.
Using Fourier transform of a special function (see \cite{RR, V,Y1,Y2}) with translation, 
we can construct a typical example of tubular vorticity $\widetilde W^{n,\zeta}$ within a band-pass  filter:
\begin{equation*}
\widetilde W^{n,\zeta}(x):=\left(0,0,\prod_{k=1}^n\left(\frac{\sin (2^{-k}x_1)}{2^{-k}x_1}\frac{\sin (2^{-k}x_2)}{2^{-k}x_2}\frac{\sin (2^{-k}x_3)}{2^{-k}x_3}\right)\sin (\zeta x_1)\right)
\end{equation*}
for some $\zeta>m:=\sum_{k=1}^n2^{-k}$ ($1/2\leq m<1$).
We will fix these parameters $n\in\mathbb{Z}_{\geq 1}$ and $\zeta\in\mathbb{R}_{>0}$ later.
Then we see that  
$\mathcal{F}[\widetilde W^{n,\zeta}]$ is smooth ($\mathcal{F}[\widetilde W^{n,\zeta}]\in C^{n-2}(\mathbb{R}^3)$) and compactly supported.
More precisely, 
\begin{equation*}
\begin{split}
&\text{supp}\,\mathcal{F}[W^{n,\zeta}]\subset\\
&
\footnotesize{
\bigg\{\xi\in\mathbb{R}^3: \sqrt{\frac{\zeta-m}{\zeta+m}}K<  \min\{|\xi_1|, |\xi_2|,|\xi_3|\}
\ \text{and}
\
\max\{|\xi_1|, |\xi_2|,|\xi_3|\}\leq \sqrt{\frac{\zeta+m}{\zeta-m}}K\bigg\}
}
\end{split}
\end{equation*}
for $K=\sqrt{\frac{\zeta+m}{\zeta-m}}(\zeta-m)$ (c.f. \eqref{LP-decomposition-2}  and \eqref{LP-decomposition}).
Then we can define a typical example of antiparallel tubular vortices $W^{n,\zeta}$ as follows:
\begin{equation*}
W^{n,\zeta}(x):=\widetilde W^{n,\zeta}(x)\chi_{[-\frac{\pi}{\zeta},\frac{\pi}{\zeta})\times[-2\pi,2\pi)^2}(x_1,x_2,x_3).
\end{equation*}
In this calculable setting, we cut off in the real space  (not in the Fourier space).
So it is rather natural to estimate 
\begin{equation*}
\frac{\int|W^{n,\zeta}-\widetilde W^{n,\zeta}|^2}{\int |\widetilde W^{n,\zeta}|^2}
\quad\text{instead of}\quad 
\frac{\int|W^{n,\zeta}-\widetilde W^{n,\zeta}|^2}{\int |W^{n,\zeta}|^2}.
\end{equation*}
 In this paper, for the sake of simplicity, we have used the characteristic function type of band-pass filter. But, rigorously, using the overlapping smoothed Littlewood-Paley decomposition may be more effective to capture antiparallel tubular vortices.
Thus it may be better to choose $\zeta$ and $m$ satisfying $\sqrt{(\zeta+m)/(\zeta-m)}>\sqrt{1.7}$.
In this appendix, we examine the case $n=2$ and $\zeta=5/4$ (in this case $m=3/4$ and $\sqrt{(\zeta+m)/(\zeta-m)}=2$).
With the aid of the numerical computation, we can estimate as the following: 
\begin{equation*}
\frac{\int|W^{n,\zeta}-\widetilde W^{n,\zeta}|^2}{\int |\widetilde W^{n,\zeta}|^2}\simeq 1-0.80\times (0.994)^2
\simeq 0.21.
\end{equation*}
From this estimate, we can naturally imagine that there exist various antiparallel tubular vortices for small $\delta$.

\vspace{0.5cm}
\noindent
{\bf Acknowledgments.}\ 
Research of TY  was partly supported by the JSPS Grants-in-Aid for Scientific
Research  17H02860, 18H01136, 18H01135 and 20H01819.
SG was partly supported by the JSPS Grants-in-Aid for Scientific Research  20H02068 and 20K20973. This work was done while TY was an associate professor at the University of Tokyo, Japan.

\bibliographystyle{amsplain}

\end{document}